\documentclass[twocolumn,showpacs,superscriptaddress,pra,floatfix]{revtex4-1}
\usepackage{graphicx}
\usepackage{amsmath}
\usepackage{amssymb}
\usepackage{latexsym}
\usepackage{array}
\usepackage{amsfonts}
\usepackage{mathrsfs}
\usepackage{color}
\usepackage{hyperref}
\usepackage{color}

\begin{document}
\title{Greenberger-Horne-Zeilinger theorem for $N$ qudits}

\author{Junghee Ryu}

\affiliation{Institute of Theoretical Physics and Astrophysics, University of Gda\'{n}sk, 80-952 Gda\'{n}sk, Poland}
\affiliation{Department of Physics, Hanyang University, Seoul 133-791, Korea}

\author{Changhyoup Lee}

\affiliation{Centre for Quantum Technologies, National University of Singapore, 3 Science Drive 2, Singapore 117543}
\affiliation{Department of Physics, Hanyang University, Seoul 133-791, Korea}

\author{Marek \.{Z}ukowski}

\affiliation{Institute of Theoretical Physics and Astrophysics, University of Gda\'{n}sk, 80-952 Gda\'{n}sk, Poland}

\author{Jinhyoung Lee}

\affiliation{Department of Physics, Hanyang University, Seoul 133-791, Korea}
\affiliation{Center for Macroscopic Quantum Control, Seoul National University, Seoul, 151-742, Korea}

%\received{\today}
\begin{abstract}
We generalize Greenberger-Horne-Zeilinger (GHZ) theorem to an arbitrary number of $D$-dimensional systems. Contrary to conventional approaches using compatible composite observables, we employ incompatible and concurrent observables, whose common eigenstate is still a generalized GHZ state. It is these concurrent observables which enable to prove a genuinely $N$-partite and $D$-dimensional GHZ theorem. Our principal idea is illustrated for a four-partite system with $D$ which is an arbitrary multiple of $3$. By extending to $N$ qudits, we show that GHZ theorem holds as long as $N$ is not divisible by all nonunit divisors of $D$, smaller than $N$.
\end{abstract}
\maketitle

\newcommand{\bra}[1]{\langle #1\vert} 
\newcommand{\ket}[1]{\vert #1\rangle} 
\newcommand{\abs}[1]{\vert#1\vert} 
\newcommand{\avg}[1]{\langle#1\rangle}
\newcommand{\braket}[2]{\langle{#1}|{#2}\rangle}
\newcommand{\commute}[2]{\left[{#1},{#2}\right]}

\newtheorem{theorem}{Theorem}

%--------------------------------------------------------------
\section{Introduction}
%--------------------------------------------------------------

The inconsistency of (local) hidden variable theories with quantum mechanics fascinates many researchers. It has been discussed in many theoretical~\cite{Bell64,*Clauser69,*Specker60,*Kochen67,*Leggett03} and experimental works~\cite{Aspect82,*Lapkiewicz11,*Groeblacher07}. Bell's theorem, one of the most profound discoveries concerning the foundations of quantum mechanics, states that any local realistic theory is incompatible with quantitative predictions of quantum mechanics. Even though Bell's theorem was studied mostly in terms of statistical inequalities, a more striking conflict, without inequalities, was also shown for a multiqubit system by Greenberger, Horne, and Zeilinger (GHZ)~\cite{GHZ89,*Pan12}. They derived an {\it all-versus-nothing} contradiction based on perfect correlations for so-called GHZ states. This leads to a direct refutation of Einstein-Podolsky-Rosen (EPR) ideas on the relation between locality and elements of reality with quantum mechanics~\cite{EPR35}.

This is a striking blow right into the very basic ideas linked with local hidden variables. After all, EPR used the concept of (local) elements of reality to support their claim that quantum mechanics is incomplete. All this can be best explained using the three particle GHZ paradox. Take a state $\ket{GHZ}=\frac{1}{\sqrt{2}}(\ket{+++}-\ket{---})$, where $\ket{\pm}$ denotes   states associated with the eigenvalues $\pm1$ of the local Pauli $\sigma_z$ operator. The
 operators $\sigma_x\otimes\sigma_x\otimes\sigma_x$,  $\sigma_x\otimes\sigma_y\otimes\sigma_y$, $\sigma_y\otimes\sigma_x\otimes\sigma_y$, and $\sigma_y\otimes\sigma_y\otimes\sigma_x$ all commute, and their eigenstate is $|GHZ\rangle$. The eigenvalues are $-1$, $1$, $1$, and $1$,
respectively, which signify perfect (GHZ)-EPR correlations. This would please any local realist. Assume that  the particles are far away from each other, and three distant independent observers can perform experiments on them, choosing at will the observables. For example, the first and the second one may choose $\sigma_x$ and their measurement results are $1$ and $-1$, respectively. In such a case, they can together  predict with certainty what would have been the result of the third observer had he or she chosen to measure also $\sigma_x$. Simply the local results must multiply to the eigenvalue of the joint observable    $\sigma_x\otimes\sigma_x\otimes\sigma_x$, and this is $-1$. Thus, the third observer, if the hypothetical case of him or her choosing to measure $\sigma_x$ really happens, must for sure get $1$. Thus, as EPR would say, this value is an {\em element of reality}, because in no way the distant choices, and obtained results can influence anything that happens at the location of the third observer (especially if measurements actions, are spatially separated events in the relativistic sense of this term, and the measurement choices are made some time after the particles are emitted from a common, say central, source). For such counterfactual reasonings one can use any of the four perfect correlation cases for the  joint measurements given above, and apply to each observer. Thus, it seems that one can ascribe elements of reality of to all local situations, no matter whether the local  observable is $\sigma_x$ or $\sigma_y$. Note that these are incommensurable. Let us denote such elements of reality, related with a single emission act of three particles, by $r_{w}^k$, where $w=x,y$ denotes the observable, and $k=1,2,3$ denotes the observer. Obviously  $r_{w}^k=\pm 1. $ For the four cases of GHZ-EPR perfect correlations one therefore must have $r_{x}^1r_{x}^2r_{x}^3=-1$, and $r_{x}^1r_{y}^2r_{y}^3=1$,  $r_{y}^1r_{x}^2r_{y}^3=1$,  $r_{y}^1r_{y}^2r_{x}^3=1$. If one multiplies these four relations side by side, one gets $1=-1$. Thus an attempt of introducing EPR elements of reality leads to a nonsense. {\em Ergo}, elements of reality are a nonsense. 
No other argument against local realism could be more striking.

Extending Bell's theorem to more complex systems such as multipartite and/or high-dimensional systems ~\cite{Mermin90c,*Werner01,*Zukowski02,*Collins02,*Laskowski04,*Son06,*James10} is important not only for a deeper understanding of foundations of quantum mechanics. It is associated with developing new applications in quantum information processing, such as quantum cryptography, secret sharing, quantum teleportation, reduction of communication complexity, quantum key distribution, and random numbers generation~\cite{Ekert91,Horodecki96,Cleve97, *Brukner04,Zukowski98,*Hillery99,*Kempe99,*Scarani01,*Barrett05,*Acin07,Pironio10}. Similarly to Bell's theorem, also all-versus-nothing tests, which we call GHZ theorem, have been generalized to higher dimensional systems. For the sake of convenience, we shall use the tuple $(N, M, D)$ to denote $N$ parties, $M$ measurements for each party, and $D$ distinct outcomes for each measurement. In Ref.~\cite{Zukowski99}, the GHZ theorem was derived for a $(D+1, 2, D)$ problem. A probabilistic but conclusive GHZ-like test was shown for $(D,2,D)$ in Ref.~\cite{Kaszlikowski02}. The $(N,2,D)$ problem for odd $N>D$ and even $D$ was studied by Cerf {\it et al.}~\cite{Cerf02a}. Lee {\it et al.} showed the GHZ theorem for more general cases, $(\mbox{odd}~N, 2, \mbox{even}~D)$, by an unconventional approach using incompatible observables~\cite{Lee06}. Recently, Tang {\it et al.} generalized GHZ theorem to the $N (\geq4)$-partite case and  even-$D$ dimensional systems with the help of GHZ graphs~\cite{Tang13}. Despite such an intensive progress in extending GHZ theorem, many cases of $N$-partite and $D$-dimensional systems remain still as open problems.

We generalize the GHZ theorem to three or higher $D$-dimensional systems. To this end, we employ concurrent composite observables which, in contrast with the standard approach,  are mutually incompatible but still have a common eigenstate, here a generalized GHZ state. They can be realized by multiport beam splitters and phase shifters, as it is shown in Refs.~\cite{Zukowski99,Lee06}. We first illustrate our principal idea with four $3d$-dimensional systems and then provide a systematic method, so as to extend it to three or higher $D$-dimensional systems. Finally, we show a GHZ-type contradiction, as long as $N$ is not divided by all nonunit divisors of $D$, smaller than $N$. Our generalization is genuinely $N$-partite {\it and} $D$-dimensional and can reproduce the previous results~\cite{GHZ89,Cerf02a,Zukowski99,Lee06}. This approach can lead to a general GHZ theorem for $N$ qudits.

%--------------------------------------------------------------
\section{Concurrent observables}{\label{sec_concurrent}}
%--------------------------------------------------------------
Some sets of observables have a common eigenstate. If a system is prepared in the eigenstate, the measurement results for such observables are concurrently appearing with certainty. Such observables are called ``concurrent''~\cite{Lee06}. For a quantum system of dimension $D (>2)$, consider two Hermitian operators $\hat{A}$ and $\hat{B}$ such that $\hat{A}=a\ket{\psi}\bra{\psi}+\hat{A}_{\psi}^{\perp}$ and $\hat{B}=b\ket{\psi}\bra{\psi}+\hat{B}_{\psi}^{\perp}$ with $\hat{A}_{\psi}^{\perp}(\hat{B}_{\psi}^{\perp}) \ket{\psi}=0$. The state $\ket{\psi}$ is then a common eigenstate of both observables as $\hat{A}(\hat{B})\ket{\psi}=a(b)\ket{\psi}$, even if $[\hat{A},\hat{B}]=[\hat{A}_{\psi}^{\perp},\hat{B}_{\psi}^{\perp} ]\neq0$~\footnote{Note that compatible observables are clearly concurrent.}.

Such concurrent observables can be constructed by the method introduced in Ref.~\cite{Lee06}. Consider a  unitary operator $\hat{U}$, which is of the form of $\hat{U}=e^{i \phi} \ket{\psi}\bra{\psi}+\hat{U}_{\psi}^{\perp}$ with $\hat{U}_{\psi}^{\perp} \ket{\psi}=0$. Here $\hat{U}_{\psi}^{\perp}$ is a unitary operator on a  space $\mathcal{H}_{\psi}^{\perp}$ which is defined by the requirement $\mathcal{H} =\mathcal{H}_{\psi} \oplus \mathcal{H}_{\psi}^{\perp}$, where $\mathcal{H}_{\psi}$ is the one-dimensional  space containing $\ket{\psi}$. Every such  unitary operator leaves the state $\ket{\psi}$ unchanged, up to  a global phase: If the  state $\ket{\psi}$  satisfies $\hat{A}\ket{\psi}=\lambda \ket{\psi}$, then all transformed operators $\hat{B}_{U}=\hat{U} \hat{A} \hat{U}^{\dagger}$ are concurrent with $\hat{A}$.

Consider $N$ qudits prepared in a generalized GHZ state $\ket{\psi}=\frac{1}{\sqrt{D}}\sum_{n=0}^{D-1} \bigotimes_{k=1}^{N} \ket{n}_{k}$, where $\ket{n}$ denotes a basis state for a qudit. This GHZ state is a common eigenstate with the unity eigenvalue of any composite observable $\hat{X}^{\otimes N} \equiv \hat{X}\otimes \hat{X} \otimes \cdots \otimes \hat{X}$ as $\hat{X}^{\otimes N} \ket{\psi}=\ket{\psi}$, where the local observable $\hat{X}$ is defined by applying quantum Fourier transformation $\hat{F}$ on a reference unitary observable $\hat{Z}=\sum_{n=0}^{D-1} \omega^n \ket{n} \bra{n}$ with $\omega=\exp({2 \pi i/D})$, that is $\hat{X}=\hat{F}\hat{Z}\hat{F}^{\dagger}$~\footnote{Here, we shall use local unitary observables $\hat{V}=\sum_{n=0}^{D-1}\omega^n \left|n\right>_{\mathrm{v}}\left<n\right|$, where $\omega=\exp(2 \pi i/D)$, in $D$-dimensional Hilbert space. The observable $\hat{V}$ is unitary, however, one can uniquely relate it with a Hermitian observable $\hat{H}$ by requiring $\hat{V}=\exp(i \hat{H})$. Therefore the complex eigenvalues of $\hat{V}$ can be associated with the measurement results, denoted by real eigenvalues of $\hat{H}$. Such a unitary representation leads to a simplification of mathematics without changing any physical results~\cite{Cerf02a, Lee06}.}. An eigenvector of $\hat{X}$ associated with the eigenvalue $\omega^n$ is given by $ \ket{n}_\mathrm{x} =\hat{F}\ket{n} = \frac{1}{\sqrt{D}} \sum_{m=0}^{D-1} \omega^{nm} \ket{m}$. With the standard basis set $\{\ket{n} \}$, the observable $\hat{X}$ is written as $\hat{X} = \sum_{n=0}^{D-1} \ket{n}\bra{n+1}$, where $\ket{n} \equiv \ket{n \mod D}$.

To construct a set of concurrent composite observables, we employ a unitary operation in the form of $\hat{U}= \bigotimes_{k=1}^{N} \hat{P}_{k} (f_k)$ with a phase shifter $\hat{P}_k=\sum_{n=0}^{D-1} \omega^{f_k(n)} \ket{n}\bra{n}$. If ``phases" $f_{k}(n)$ satisfy a condition,
\begin{equation}
\sum_{k=1}^{N} f_{k} (n) \equiv 0 \mod D,
\label{invariant_condition}
\end{equation}
for each $n$, then the unitary operator $\hat{U}$ leaves the GHZ state $\ket{\psi}$ invariant. This simple invariance condition enables one to construct a large number of concurrent observables which have a common eigenstate of the generalized GHZ state. 

Let us apply the unitary operation $\hat{U}$ with the phases $f_{k}(n)=\alpha_{k} n$ with rational numbers $\alpha_{k}$ to $\hat{X}^{\otimes N}$. If the phases $f_{k} (n)$ satisfy the invariance condition~(\ref{invariant_condition}), the transformed observable $\hat{U} \hat{X}^{\otimes N} \hat{U}^{\dagger}=\hat{X}(\alpha_1)\otimes \hat{X}(\alpha_2)\otimes \cdots \otimes\hat{X}(\alpha_N)$ is concurrent with $\hat{X}^{\otimes N}$, i.e., $\hat{U} \hat{X}^{\otimes N} \hat{U}^{\dagger} \ket{\psi}=\ket{\psi}$. For each eigenvalue $\omega^{n}$, the eigenvector of the local observable $\hat{X}({\alpha})$ is given by applying the phase shifter $\hat{P}$ on $\ket{n}_{\mathrm{x}}$ as $\ket{n}_{\alpha} = \hat{P}\ket{n}_\mathrm{x}= \frac{1}{\sqrt{D}}\sum_{m=0}^{D-1} \omega^{(n+\alpha)m} \ket{m}$.
The observable $\hat{X}({\alpha})$ can be written in the standard basis set $\{\ket{n} \}$ as
\begin{equation}
\hat{X}({\alpha})=\omega^{-\alpha} \left( \sum_{n=0}^{D-2} \ket{n}\bra{n+1} + \omega^{\alpha D} \ket{D-1}\bra{0} \right).
\label{obs_y}
\end{equation}
Note that if $\alpha$ is an integer, the measurement basis set $\{ \ket{n}_{\alpha}\}$ of $\hat{X}(\alpha)$ will be the same as $\{\ket{n}_\mathrm{x} \}$ of $\hat{X}$ except the ordering, i.e., $\ket{n}_{\alpha} = \ket{n + \alpha}_\mathrm{x} $. Thus, $\hat{X}(\alpha) = \omega^{-\alpha} \hat{X}$. That is, the observable $\hat{X}(\alpha)$ is equivalent to $\hat{X}$, up to a  phase factor  $\omega^{-\alpha}$.  Let $\ket{n}_{\alpha}$ be the eigenstate of $\hat{X}(\alpha)$ associated with eigenvalue $\omega^n$, and $\ket{m}_{\beta}$ the eigenstate of $\hat{X}(\beta)$ associated with $\omega^m$. If and only if $\alpha$ differs from $\beta$ by an integer, then two measurement bases satisfy $\abs{{}_{\alpha}\bra{n} m \rangle_{\beta}}^2 = \delta_{D} (\gamma)$,
where $\gamma=m-n+\beta-\alpha$. Here $\delta_{D}(\gamma)=1$ if $\gamma$ is congruent to zero modulo $D$ and otherwise $\delta_{D}(\gamma)=0$. That is, if $\beta - \alpha$ is not an integer, two local observables $\hat{X}(\alpha)$ and $\hat{X}(\beta)$ are inequivalent. 

%\begin{figure}[t]
%\includegraphics[width=4cm]{figure1}
% \caption{The invariant condition for $4$-partite system. Each party applies a local phase shifter $\hat{P}_k$ with phase $f_{k} (n)$ to the generalized GHZ state $\ket{\psi}$. If the sum of the local phases $f_{k} (n)$ is congruent to zero modulo $D$, i.e., $f_1 (n)+f_2 (n)+f_3 (n)+f_4 (n) \equiv 0 \mod D$, then the state $\ket{\psi}$ transformed by the phase shifters is unchanged.}
% \label{fig_phase shift}
%\end{figure}

%--------------------------------------------------------------
\section{Generalized GHZ theorem}{\label{sec_4partite}}
%--------------------------------------------------------------
\subsection{Four-qudit system}
We first illustrate our idea by considering a four-qudit system. Already this case goes significantly beyond the previous studies~\cite{Zukowski99, Cerf02a, Lee06, Tang13}. Take a four-qudit GHZ state $\ket{\psi}=\frac{1}{\sqrt{D}}\sum_{n=0}^{D-1} \ket{n,n,n,n}$ (here $D$ is assumed to be an integral  multiple of 3). 
%\begin{equation}
%\ket{\psi}=\frac{1}{\sqrt{D}}\sum_{n=0}^{D-1} \ket{n,n,n,n}.
%\label{4_ghz_state}
%\end{equation}
The qudits are distributed to four sufficiently separated parties. Each party performs one of two nondegenerate local measurements on his or her qudit, each of which produces distinguishable $D$ outcomes. We represent the measurement for the $k$-th party by $\hat{M}_k$, and the eigenvalues of the observables are of the form $\omega^{m_k}$, where $m_k$ is an integer. One can denote a joint probability that each party obtains the result $\omega^{m_k}$ by $\mathcal{P}(m_1,m_2,m_3,m_4)$, and define a correlation function $E_{\mathrm{QM}} (M_1,M_2,M_3,M_4)=\bra{\psi}\hat{M}_1\otimes\hat{M}_2\otimes\hat{M}_3\otimes\hat{M}_4\ket{\psi}$, equivalently the quantum average of products of the measurement results: $\sum_{m_{1}=0}^{D-1}\cdots\sum_{m_{4}=0}^{D-1} \omega^{\sum_{i=1}^{4} m_i} \mathcal{P} (m_1,m_2,m_3,m_4)$.
%\begin{eqnarray}
%\sum_{m_{1}=0}^{D-1}\cdots\sum_{m_{4}=0}^{D-1} \omega^{\sum_{i=1}^{4} m_i} \mathcal{P} (m_1,m_2,m_3,m_4).
%\label{eq:correlation}
%\end{eqnarray}
When all measurements are $\hat{X}$, that is, each $\hat{M}_i=\hat{X}$, since $\hat{X} \otimes \hat{X} \otimes \hat{X} \otimes \hat{X} \ket{\psi} = \ket{\psi}$, one has $ E_{\mathrm{QM}} (X,X,X,X)=1$. 
This implies that we have a perfect GHZ correlation. If arbitrary three parties know their own outcomes $\omega^{x_i}$ of measurements $\hat{X}$, then they can predict with certainty the remaining party's outcome. We will denote such a perfect correlation by
\begin{equation}
C_{\mathrm{QM}} (x_1 + x_2 + x_3 + x_4 \equiv 0 ).
\label{eq:perfect_correlation}
\end{equation}
The sum is taken modulo $D$; such a convention is used below in all formulas.

Let us construct concurrent composite observables from the  observable $\hat{v}_0 = \hat{X}^{\otimes 4}$, by applying a unitary operator $\hat{U}_{1}=\hat{P}_{1}\otimes\hat{P}_{2}^{\otimes3}$, with the phase shifters $\hat{P}_k$ of phases $f_{1}(n)=(D-1)n$ and $f_{2}(n)=n/3$. One of the new observables is $\hat{v}_{1} \equiv \hat{U}_{1} \hat{v}_{0} \hat{U}_{1}^{\dagger} = \hat{X}(D-1) \otimes \hat{X}(1/3)^{\otimes 3}$. The phases $f_k (n)$ satisfy the invariance condition~(\ref{invariant_condition}), $f_1 (n) + 3 f_2 (n) \equiv 0 \mod D$. Thus,  the observable $\hat{v}_1$  has $|\psi\rangle $  as its eigenstate with  eigenvalue $1$. By Eq.~(\ref{obs_y}) one has $\hat{X}(D-1) =\omega \hat{X}(0)$, and $\hat{Y} \equiv \hat{X}(1/3)$ is given by
\begin{equation}
\hat{Y}=\omega^{-1/3} \left( \sum_{n=0}^{D-2} \ket{n}\bra{n+1} + \omega^{D/3} \ket{D-1}\bra{0} \right).
\label{4_obs_y}
\end{equation}
The observable $\hat{Y}=\hat{X}(1/3)$ is not equivalent to $\hat{X}=\hat{X}(0)$ [see the explanation below Eq.~(\ref{obs_y}]. The other three concurrent observables are obtained by $\hat{v}_l \equiv \hat{U}_{l} \hat{v}_0 \hat{U}_l^{\dagger},~l\in \{2,3,4\}$, where the unitary operators $\hat{U}_l$ are composed of the cyclic permutations of the local phase shifters: $\hat{U}_{2}=\hat{P}_{2}\otimes\hat{P}_{1}\otimes\hat{P}_{2}\otimes\hat{P}_{2}, \hat{U}_{3}=\hat{P}_{2}\otimes\hat{P}_{2}\otimes\hat{P}_{1}\otimes\hat{P}_{2}$, and $\hat{U}_{4}=\hat{P}_{2}\otimes\hat{P}_{2}\otimes\hat{P}_{2}\otimes\hat{P}_{1}$. The perfect  correlations for $\hat{v}_{i}$ $(i=1,\dots,4)$ are, respectively, given by
\begin{eqnarray}
&C_{\mathrm{QM}} (x_1 + y_2 + y_3 + y_4 \equiv -1) &, \nonumber \\
&C_{\mathrm{QM}} (y_1 + x_2 + y_3 + y_4 \equiv -1) &, \nonumber \\
&C_{\mathrm{QM}} (y_1 + y_2 + x_3 + y_4 \equiv -1) &, \nonumber \\
&C_{\mathrm{QM}} (y_1 + y_2 + y_3 + x_4 \equiv -1) &,
\label{prob_QM}
\end{eqnarray}
where $\omega^{y_i}$ is an outcome of measurement $\hat{Y}$ for the $i$-th party. 

Local realistic theories assume that the outcomes of the measurements are predetermined, before the actual measurements.
This implies that the values of local realistic predictions for the correlations (\ref{prob_QM}), for each experimental run, must satisfy:
\begin{eqnarray}
& \omega^{x_1}\omega^{y_2}\omega^{ y_3}\omega^{y_4} = \omega^{-1},& \nonumber \\
& \omega^{y_1}\omega^{x_2}\omega^{ y_3}\omega^{y_4} =  \omega^{-1},& \nonumber \\
& \omega^{y_1}\omega^{y_2}\omega^{ x_3}\omega^{y_4} =  \omega^{-1},& \nonumber \\
& \omega^{y_1}\omega^{y_2}\omega^{ y_3}\omega^{x_4} =  \omega^{-1}.&
\label{eq_LHV2}
\end{eqnarray}
If local realism is also to reproduce quantum perfect correlations for the 
case of $\hat{v}_0 = \hat{X} ^{\otimes4}$,
one must have 
%\begin{equation}
%E_{\mathrm{LR}} (  x_1, x_2, x_3, x_4   ) = 1,
%\label{constraint_eq}
%\end{equation}
for each run,
\begin{eqnarray}
& \omega^{x_1}\omega^{x_2}\omega^{ x_3}\omega^{x_4} =1.&  \label{eq_LHV2B}
\end{eqnarray}
However, one sees that this is possible only provided $\omega^{  -3 \sum_{i=1}^{4} y_{i}-4} =1 $ if one multiplies side-by-side all equations (\ref{eq_LHV2}). As $\omega=\exp(2\pi i/D)
$, 
if $D=3d$ where $d$ is an integer, one can use the fact that an elementary algebra shows that there is no integer solution of the equation $3{\bf y} + 4 \equiv 0 \mod D$. Thus local realistic correlation (\ref{eq_LHV2B}) is impossible.

It is worth noting that the approach with  concurrent observables relaxes the restrictions of early studies requiring {\it compatible} observables. This enables one to generalize GHZ contradictions beyond the case $N>D$ studied in Refs.~\cite{Zukowski99,Cerf02a}.

Note, that to prove the four-partite GHZ contradiction, we chose the local dimension $D$ and the number of the observables $\hat{Y}$'s in the considered correlation functions, $N_2$, such that the greatest common divisor (gcd) of $D$ and $N_2$, here $\mathrm{gcd}(D,N_2)=3$, does not divide the number of parties $N=4$,  or equivalently the number $N_1$ of the observables $\hat{X}$ (here equal to $1$). This mathematical property plays a central role in the generalization of the GHZ contradiction to an arbitrary number of parties.

%The unitary observables $\hat{X}$ and $\hat{Y}$  can be realized by using optical devices such as multiport beam splitters and phase shifters (see, \cite{Reck94,*Zukowski97}). 

%--------------------------------------------------------------
\subsection{Extending to $N$ qudits system}{\label{sec_extend}}
%--------------------------------------------------------------

We extend our approach into a general case of $N$ qudits, $N\geq3$, such that $N$ is nondivisible by any nonunit divisor of $D$, smaller than $N$.
To this end, we use a set of $(N+1)$ concurrent observables given by $\hat{v}_{0}=\hat{X}^{\otimes N}$ and $N$ observables of the following forms:
 $\hat{v}_{1}=\omega \hat{X}^{\otimes N_1} \otimes \hat{Y}^{\otimes N_2}$
and
 $\hat{v}_{k}= \hat{Y}^{\otimes  k-1}\otimes
\omega  \hat{X}^{\otimes N_1} \otimes \hat{Y}^{\otimes  N_2-k+1}$
for $k=2, \dots, N_2 +1$
and finally
 $\hat{v}_{k}=  \hat{X}^{\otimes k-N_2 -1} \otimes \hat{Y}^{\otimes  N_2}\otimes
\omega  \hat{X}^{\otimes N-k+1} $
for $k=N_2+2,\dots, N$.

The composite observable $\hat{v}_1$ is obtained by a unitary transformation,  $\hat{U}_{1}=\hat{P}_{1}\otimes \openone^{\otimes N_1 -1} \otimes \hat{P}_{2}^{\otimes N_2}$,  of the observable $\hat{v}_{0}$, i.e., $\hat{v}_{1}=\hat{U}_{1}\hat{v}_{0}\hat{U}_{1}^{\dagger}$, with the phase shifters $\hat{P}_1$ and $\hat{P}_2$ of  phases $f_{1} (n)=(D-1)n$ and $f_{2}(n)=n/N_2$, respectively. The local observables $\hat{X}=\hat{X}(0)$ and $\hat{Y}=\hat{X}(1/N_{2})$ are given by Eq.~(\ref{obs_y}). Likewise, we obtain the other concurrent observables $\hat{v}_{l}~(2 \leq l \leq N)$ by cyclic permutations in the unitary operator $\hat{U}_{1}$, as it was done for the four-partite case.
The phases satisfy the invariance condition~(\ref{invariant_condition}) as $f_1 (n) + N_2 f_2 (n) \equiv 0 \mod D$. Thus, the $N$-partite generalized GHZ state $\ket{\psi}=\frac{1}{\sqrt{D}}\sum_{n=0}^{D-1} \bigotimes_{k=1}^{N} \ket{n}_{k}$ is a common eigenstate of all the $(N+1)$ concurrent observables $\hat{v}_l$ $(l=0,\dots,N)$, with the same eigenvalue $1$. This leads to the following values of correlation functions (for later convenience we use party indices $i=1, \dots, N$, which will later on allow us to get a more concise notation in formulas). If all local observables are $\hat{X}$, that is, for global $\hat{v}_0$, one has $E_{\mathrm{QM}} (X,X,\dots,X) = 1$. Thus we have a perfect correlation which can be denoted, in the way introduced earlier, as $C_{\mathrm{QM}} (\sum_{i=1}^{N}x_i \equiv 0)$. For $\hat{v}_k$, where $k=1,2,\dots,N$, one has perfect correlations of the following forms: for $k=1$, $C_{\mathrm{QM}} ( \sum_{i=1}^{N_1}x_i + \sum_{i=N_1+1}^{N}y_{i} \equiv -1)$,
%\begin{eqnarray}
%C_{\mathrm{QM}} ( \sum_{i=1}^{N_1}x_i + \sum_{i=N_1+1}^{N}y_{i} \equiv -1) ,  \\ \nonumber
%\end{eqnarray}
and for $k=2,\dots,N_2 +1$, $C_{\mathrm{QM}} ( \sum_{i=1}^{k-1}y_i + \sum_{i=k}^{N_1+k-1}x_{i} + \sum_{i=N_1+k}^{N}y_{i} \equiv -1)$,
%\begin{eqnarray}
%C_{\mathrm{QM}} ( \sum_{i=1}^{k-1}y_i + \sum_{i=k}^{N_1+k-1}x_{i} + \sum_{i=N_1+k}^{N}y_{i} \equiv -1) ,  \\ \nonumber
%\end{eqnarray}
and finally for $k=N_2+2,\dots, N$, $C_{\mathrm{QM}} ( \sum_{i=1}^{k-N_2 -1}x_i + \sum_{i=k-N_2}^{k-1}y_{i} + \sum_{i=k}^{N}x_{i} \equiv -1)$.
%\begin{eqnarray}
%C_{\mathrm{QM}} ( \sum_{i=1}^{k-N_2 -1}x_i + \sum_{i=k-N_2}^{k-1}y_{i} + \sum_{i=k}^{N}x_{i} \equiv -1) .  \\ \nonumber
%\label{prob_QM2}
%\end{eqnarray}

Following similar arguments as in  the case of the four-partite GHZ contradiction, we obtain the following condition for  the local realistic correlation function for the composite observable $\hat{v}_{0}=\hat{X}^{\otimes N}$, to have value equal to the quantum prediction, that is, 1. It reads (modulo $D$)
\begin{equation} N_1 \sum_{i=1}^{N} x_{i} \equiv -N_2 \sum_{i=1}^{N} y_{i} - N\equiv 0 .
\label{general_LR_condition}
\end{equation}
However,  if $N_2$ is an integral multiple of $g$ but $N$ cannot be divided by $g$, then there are no solutions of ${\bf y}=\sum_{i} y_{i}$ to the equation $N_2 {\bf y} + N \equiv 0 \mod D$. The greatest common divisor of $N_2$ and $D$ is an integral multiple of $g$, i.e., gcd($N_2, D$)=$kg$ for some positive integer $k$ but $kg$ cannot divide $N$ as $N$ is not an  integer multiple of $g$. Thus we have a contradiction with the quantum prediction.

In order to show a GHZ contradiction for $N$-partite and $D$-dimensional system, we choose that (a) $D=dg$, (b) $N_2 = \eta g$, where $d$ and $\eta$ are positive integers, and (c) $N$ cannot be divided by $g$. Choosing the integer $g$, a nonunit divisor  $D$, plays a crucial role. For example, consider {\em four} six-dimensional systems. The nonunit divisors $g$, smaller than $N=4$, are $2$ and $3$. If we choose $g=2$, then we are unable to see any four-partite GHZ contradiction as the greatest common divisor (gcd) of $N_2$ and $D$, $\mathrm{gcd}(N_2 = 2, D=6)=2$, divides $N=4$. On the other hand, if we choose $g=3$, the four-partite GHZ contradiction can be proved as $\mathrm{gcd}(N_2 =3 , D=6)=3$ and $g=3$ does not divide $N=4$. This is a specific example of a GHZ contradiction for a $(4,2,3d)$ problem. As a consequence, we conclude that one is always able to prove the GHZ contradiction for the $N(\geq 3)$-partite and $D(\geq 2)$-dimensional systems as long as $N$ cannot be divided by all nonunit divisors of $D$.

%--------------------------------------------------------------
%\subsection{Comparison with previous works}
%--------------------------------------------------------------

For appropriate values of $N$ and $D$, our approach reproduces the previous works~\cite{Cerf02a,Zukowski99,Lee06}. A GHZ contradiction for $(D+1)$ qudits shown in Ref.~\cite{Zukowski99} is reproduced by choosing $N_1=1$ and $N_2=D$ in our method, and noticing the fact that $N_1=1$ is indivisible by  $D=\mathrm{gcd} (N_2 =D, D)$. The case of $(\mathrm{odd}~N, 2, \mathrm{even}~D)$ studied in Refs.~\cite{Cerf02a,Lee06} can be also proved by choosing a nonunit divisor $g=2$ of $D$ and an arbitrary odd integer $N_1$. One can also easily check that if $N_2 = D=2$ and $N_1 =1$, then our contradiction is reduced to the original GHZ theorem~\cite{GHZ89}.

%--------------------------------------------------------------
\subsection{Genuinely $N$-partite $D$-dimensional case}
%--------------------------------------------------------------

The GHZ theorem for the two-dimensional systems seems to be fully understood. However,  for more complex systems this is not so. Cerf {\it et al.} suggested a criterion for  a genuinely $N$-partite ($D$-dimensional) GHZ contradiction~\cite{Cerf02a}. It arises only for the given full $N$-partite ($D$-dimensional) system, but not for any $n(<N)$-partite subset, or for an effectively lower dimensionality of  the involved observables. For example, the three-qubit classic GHZ theorem can be put as the theorem for three qutrits, and specific entangled GHZ  states involving only two-dimensional subspaces for each qutrit.
 
The GHZ contradiction we show here  is a genuinely $N$-partite one, as  it is constructed using a set of composite observables composed of cyclic permutations. Let us explain this with the  four-partite GHZ contradiction, where we used the five concurrent composite observables: $\hat{X} \otimes \hat{X} \otimes \hat{X} \otimes \hat{X}$, $\hat{X} \otimes \hat{Y} \otimes \hat{Y} \otimes \hat{Y}$, $\hat{Y} \otimes \hat{X} \otimes \hat{Y} \otimes \hat{Y}$, $\hat{Y} \otimes \hat{Y} \otimes \hat{X} \otimes \hat{Y}$, and $\hat{Y} \otimes \hat{Y} \otimes \hat{Y} \otimes \hat{X}$. In such circumstances, if we eliminate one of the parties,  we are unable to show a GHZ contradiction with the remaining  observables. The four-partite GHZ state is no longer their common eigenstate. Similar argument can be put forward in the case of our $N$-partite theorem. 

The genuine $D$ dimensionality is reflected by the fact that the operators are undecomposable to a direct sum of any subdimensional observables~\cite{Lee06}. In other words, if two local observables $\hat{X}$ and $\hat{Y}$ can be simultaneously block diagonalized by some similarity transformation $\hat{S}$ such that $\hat{S}\hat{X}\hat{S}^{\dagger}=\hat{X}_1 \oplus \cdots \oplus \hat{X}_K$ and $\hat{S}\hat{Y}\hat{S}^{\dagger}=\hat{Y}_1 \oplus \cdots \oplus \hat{Y}_K$, then there exist some eigenstates $\ket{n}_{\alpha}$ of $\hat{X}$ and $\ket{m}_{\beta}$ of $\hat{Y}$ such that ${}_{\alpha}\bra{n} \hat{S}^{\dagger} \hat{S} \ket{m}_{\beta}=0$ and one can find a sub-dimensional GHZ contradiction. However, there are no such eigenstates in our method because for every $n$ and $m$, $\abs{{}_{\alpha}\langle n \ket{m}_{\beta}}^2 = \frac{\sin^2 (\pi \xi)}{D^2 \sin^2 \left[(\pi/D) \xi \right]} > 0$,
%\begin{equation}
%\abs{{}_{\alpha}\langle n \ket{m}_{\beta}}^2 = \frac{\sin^2 (\pi \xi)}{D^2 \sin^2 \left[(\pi/D) \xi \right]} > 0,
%\label{condition_compl}
%\end{equation} 
where $\xi=m-n+\beta-\alpha$. 
As, the local observables $\hat{X}=\hat{X}(\alpha)$ and $\hat{Y}=\hat{X}(\beta)$ are such that $\beta - \alpha$ is not an  integer,
$\xi$ is a nonintegral rational number.  Thus,  our GHZ contradiction is genuinely $D$ dimensional.

%One might try to generalize our method to more than two measurement settings. However this cannot be a genuinely multi-setting case. Consider a $(5,3,2)$ problem, where six concurrent observables are required; $\hat{v}_0 = \hat{X} \otimes \hat{X} \otimes \hat{X} \otimes \hat{X} \otimes \hat{X}$, $\hat{v}_1 = \omega \hat{X} \otimes \hat{Y} \otimes \hat{Y} \otimes \hat{Z} \otimes \hat{Z}$, and $\hat{v}_{l}$  formed by cyclic permutations of observables in $\hat{v}_1 (2\leq l \leq 5)$. The local observables are given by Eq.~(\ref{obs_y}) with the $f_1 (n) = (D-1)n$, $f_2 (n)=n/8$, and $f_3 (n) = 3n/8$, respectively. One can show a GHZ contradiction with the GHZ state $(\left |0,0,0,0,0 \right> + \left |1,1,1,1,1 \right>)/\sqrt{2}$. However, it is not a genuine $3$-setting one. The contradiction also arises for observables, composed of two local measurements $\hat{X}$ and $\hat{Y}$ of a similar kind as above. Thus problem whether there are  genuinely multi-setting GHZ contradictions is still open.

%--------------------------------------------------------------
\section{Summary}
%--------------------------------------------------------------

We construct a generalized GHZ contradiction for  {\it multipartite} and {\it high}-dimensional systems. The GHZ theorem holds as long as $N$ is not divisible by all nonunit divisors of $D$, smaller than $N$. We also demonstrate that our formulation of a generalized GHZ contradiction is genuinely $N$ partite {\it and} $D$ dimensional. For this purpose, we employ concurrent composite observables, which have a generalized GHZ state as a common eigenstate (even though these observables are {\it incompatible}). Our approach, by using concurrent observables, enables us to find a broader class of GHZ contradictions. There remain still more possibilities for constructing concurrent observables, which may help in further extension of the GHZ theorem. We hope that our approach of concurrent observables would be useful in the search of other kinds of quantum correlations, which are impossible classically.

%As a result, we  showed a $4$-partite GHZ contradiction for $3d$-dimensional systems, where $d$ is an arbitrary integer.  We extended the method to $N$-partite situation with $D$-dimensional systems. It turned out that  non-unit divisors of $D$ plays a crucial  role to show the GHZ contradiction. We were always able to show a contradiction as long as $N$ cannot be divided by all non-unit divisors  of $D$, smaller than $N$. We also showed that our formulation of a generalized GHZ contradiction is genuinely $N$-partite {\it and} $D$-dimensional.

\acknowledgements
We thank \v{C}. Brukner for discussions. The work was supported by the National Research Foundation of Korea (NRF) grant funded by the Korea government (MEST) (Grants No. 2010-0015059 and No. 2010-0018295). J.R. and M.\.{Z}. are supported by the Foundation for Polish Science TEAM project cofinanced by the EU European Regional Development Fund and a NCBiR-CHIST-ERA Project QUASAR. C.L. is supported by the National Research Foundation and Ministry of Education, Singapore.

\bibliography{../../reference}

\begin{thebibliography}{36}%
\makeatletter
\providecommand \@ifxundefined [1]{%
 \@ifx{#1\undefined}
}%
\providecommand \@ifnum [1]{%
 \ifnum #1\expandafter \@firstoftwo
 \else \expandafter \@secondoftwo
 \fi
}%
\providecommand \@ifx [1]{%
 \ifx #1\expandafter \@firstoftwo
 \else \expandafter \@secondoftwo
 \fi
}%
\providecommand \natexlab [1]{#1}%
\providecommand \enquote  [1]{``#1''}%
\providecommand \bibnamefont  [1]{#1}%
\providecommand \bibfnamefont [1]{#1}%
\providecommand \citenamefont [1]{#1}%
\providecommand \href@noop [0]{\@secondoftwo}%
\providecommand \href [0]{\begingroup \@sanitize@url \@href}%
\providecommand \@href[1]{\@@startlink{#1}\@@href}%
\providecommand \@@href[1]{\endgroup#1\@@endlink}%
\providecommand \@sanitize@url [0]{\catcode `\\12\catcode `\$12\catcode
  `\&12\catcode `\#12\catcode `\^12\catcode `\_12\catcode `\%12\relax}%
\providecommand \@@startlink[1]{}%
\providecommand \@@endlink[0]{}%
\providecommand \url  [0]{\begingroup\@sanitize@url \@url }%
\providecommand \@url [1]{\endgroup\@href {#1}{\urlprefix }}%
\providecommand \urlprefix  [0]{URL }%
\providecommand \Eprint [0]{\href }%
\providecommand \doibase [0]{http://dx.doi.org/}%
\providecommand \selectlanguage [0]{\@gobble}%
\providecommand \bibinfo  [0]{\@secondoftwo}%
\providecommand \bibfield  [0]{\@secondoftwo}%
\providecommand \translation [1]{[#1]}%
\providecommand \BibitemOpen [0]{}%
\providecommand \bibitemStop [0]{}%
\providecommand \bibitemNoStop [0]{.\EOS\space}%
\providecommand \EOS [0]{\spacefactor3000\relax}%
\providecommand \BibitemShut  [1]{\csname bibitem#1\endcsname}%
\let\auto@bib@innerbib\@empty
%</preamble>
\bibitem [{\citenamefont {Bell}(1964)}]{Bell64}%
  \BibitemOpen
  \bibfield  {author} {\bibinfo {author} {\bibfnamefont {J.~S.}\ \bibnamefont
  {Bell}},\ }\href@noop {} {\bibfield  {journal} {\bibinfo  {journal}
  {Physics}\ }\textbf {\bibinfo {volume} {1}},\ \bibinfo {pages} {195}
  (\bibinfo {year} {1964})}\BibitemShut {NoStop}%
\bibitem [{\citenamefont {Clauser}\ \emph {et~al.}(1969)\citenamefont
  {Clauser}, \citenamefont {Horne}, \citenamefont {Shimony},\ and\
  \citenamefont {Holt}}]{Clauser69}%
  \BibitemOpen
  \bibfield  {author} {\bibinfo {author} {\bibfnamefont {J.~F.}\ \bibnamefont
  {Clauser}}, \bibinfo {author} {\bibfnamefont {M.~A.}\ \bibnamefont {Horne}},
  \bibinfo {author} {\bibfnamefont {A.}~\bibnamefont {Shimony}}, \ and\
  \bibinfo {author} {\bibfnamefont {R.~A.}\ \bibnamefont {Holt}},\ }\href@noop
  {} {\bibfield  {journal} {\bibinfo  {journal} {Phys. Rev. Lett.}\ }\textbf
  {\bibinfo {volume} {23}},\ \bibinfo {pages} {880} (\bibinfo {year}
  {1969})}\BibitemShut {NoStop}%
\bibitem [{\citenamefont {Specker}(1960)}]{Specker60}%
  \BibitemOpen
  \bibfield  {author} {\bibinfo {author} {\bibfnamefont {E.~P.}\ \bibnamefont
  {Specker}},\ }\href@noop {} {\bibfield  {journal} {\bibinfo  {journal}
  {Dialectica}\ }\textbf {\bibinfo {volume} {14}},\ \bibinfo {pages} {239}
  (\bibinfo {year} {1960})}\BibitemShut {NoStop}%
\bibitem [{\citenamefont {Kochen}\ and\ \citenamefont
  {Specker}(1967)}]{Kochen67}%
  \BibitemOpen
  \bibfield  {author} {\bibinfo {author} {\bibfnamefont {S.}~\bibnamefont
  {Kochen}}\ and\ \bibinfo {author} {\bibfnamefont {E.~P.}\ \bibnamefont
  {Specker}},\ }\href@noop {} {\bibfield  {journal} {\bibinfo  {journal} {J.
  Math. Mech.}\ }\textbf {\bibinfo {volume} {17}},\ \bibinfo {pages} {59}
  (\bibinfo {year} {1967})}\BibitemShut {NoStop}%
\bibitem [{\citenamefont {Leggett}(2003)}]{Leggett03}%
  \BibitemOpen
  \bibfield  {author} {\bibinfo {author} {\bibfnamefont {A.}~\bibnamefont
  {Leggett}},\ }\href@noop {} {\bibfield  {journal} {\bibinfo  {journal}
  {Found. Phys.}\ }\textbf {\bibinfo {volume} {33}},\ \bibinfo {pages} {1469}
  (\bibinfo {year} {2003})}\BibitemShut {NoStop}%
\bibitem [{\citenamefont {Aspect}\ \emph {et~al.}(1982)\citenamefont {Aspect},
  \citenamefont {Grangier},\ and\ \citenamefont {Roger}}]{Aspect82}%
  \BibitemOpen
  \bibfield  {author} {\bibinfo {author} {\bibfnamefont {A.}~\bibnamefont
  {Aspect}}, \bibinfo {author} {\bibfnamefont {P.}~\bibnamefont {Grangier}}, \
  and\ \bibinfo {author} {\bibfnamefont {G.}~\bibnamefont {Roger}},\
  }\href@noop {} {\bibfield  {journal} {\bibinfo  {journal} {Phys. Rev. Lett.}\
  }\textbf {\bibinfo {volume} {49}},\ \bibinfo {pages} {91} (\bibinfo {year}
  {1982})}\BibitemShut {NoStop}%
\bibitem [{\citenamefont {Lapkiewicz}\ \emph {et~al.}(2011)\citenamefont
  {Lapkiewicz}, \citenamefont {Li}, \citenamefont {Schaeff}, \citenamefont
  {Langford}, \citenamefont {Ramelow}, \citenamefont {Wie\'{s}niak},\ and\
  \citenamefont {Zeilinger}}]{Lapkiewicz11}%
  \BibitemOpen
  \bibfield  {author} {\bibinfo {author} {\bibfnamefont {R.}~\bibnamefont
  {Lapkiewicz}}, \bibinfo {author} {\bibfnamefont {P.}~\bibnamefont {Li}},
  \bibinfo {author} {\bibfnamefont {C.}~\bibnamefont {Schaeff}}, \bibinfo
  {author} {\bibfnamefont {N.~K.}\ \bibnamefont {Langford}}, \bibinfo {author}
  {\bibfnamefont {S.}~\bibnamefont {Ramelow}}, \bibinfo {author} {\bibfnamefont
  {M.}~\bibnamefont {Wie\'{s}niak}}, \ and\ \bibinfo {author} {\bibfnamefont
  {A.}~\bibnamefont {Zeilinger}},\ }\href@noop {} {\bibfield  {journal}
  {\bibinfo  {journal} {Nature}\ }\textbf {\bibinfo {volume} {474}},\ \bibinfo
  {pages} {490} (\bibinfo {year} {2011})}\BibitemShut {NoStop}%
\bibitem [{\citenamefont {Gr\"{o}blacher}\ \emph {et~al.}(2007)\citenamefont
  {Gr\"{o}blacher}, \citenamefont {Paterek}, \citenamefont {Kaltenbaek},
  \citenamefont {\v{C}. Brukner}, \citenamefont {\.{Z}ukowski}, \citenamefont
  {Aspelmeyer},\ and\ \citenamefont {Zeilinger}}]{Groeblacher07}%
  \BibitemOpen
  \bibfield  {author} {\bibinfo {author} {\bibfnamefont {S.}~\bibnamefont
  {Gr\"{o}blacher}}, \bibinfo {author} {\bibfnamefont {T.}~\bibnamefont
  {Paterek}}, \bibinfo {author} {\bibfnamefont {R.}~\bibnamefont {Kaltenbaek}},
  \bibinfo {author} {\bibnamefont {\v{C}. Brukner}}, \bibinfo {author}
  {\bibfnamefont {M.}~\bibnamefont {\.{Z}ukowski}}, \bibinfo {author}
  {\bibfnamefont {M.}~\bibnamefont {Aspelmeyer}}, \ and\ \bibinfo {author}
  {\bibfnamefont {A.}~\bibnamefont {Zeilinger}},\ }\href@noop {} {\bibfield
  {journal} {\bibinfo  {journal} {Nature}\ }\textbf {\bibinfo {volume} {446}},\
  \bibinfo {pages} {871} (\bibinfo {year} {2007})}\BibitemShut {NoStop}%
\bibitem [{\citenamefont {Greenberger}\ \emph {et~al.}(1989)\citenamefont
  {Greenberger}, \citenamefont {Horne},\ and\ \citenamefont
  {Zeilinger}}]{GHZ89}%
  \BibitemOpen
  \bibfield  {author} {\bibinfo {author} {\bibfnamefont {D.~M.}\ \bibnamefont
  {Greenberger}}, \bibinfo {author} {\bibfnamefont {M.~A.}\ \bibnamefont
  {Horne}}, \ and\ \bibinfo {author} {\bibfnamefont {A.}~\bibnamefont
  {Zeilinger}},\ }in\ \href@noop {} {\emph {\bibinfo {booktitle} {Bell's
  Theorem, Quantum Theory, and Conceptions of the Universe}}},\ \bibinfo
  {editor} {edited by\ \bibinfo {editor} {\bibfnamefont {M.}~\bibnamefont
  {Kafatos}}}\ (\bibinfo  {publisher} {Kluwer, Dordrecht},\ \bibinfo {year}
  {1989})\BibitemShut {NoStop}%
\bibitem [{\citenamefont {Pan}\ \emph {et~al.}(2012)\citenamefont {Pan},
  \citenamefont {Chen}, \citenamefont {Lu}, \citenamefont {Weinfurter},
  \citenamefont {Zeilinger},\ and\ \citenamefont {\.{Z}ukowski}}]{Pan12}%
  \BibitemOpen
  \bibfield  {author} {\bibinfo {author} {\bibfnamefont {J.-W.}\ \bibnamefont
  {Pan}}, \bibinfo {author} {\bibfnamefont {Z.-B.}\ \bibnamefont {Chen}},
  \bibinfo {author} {\bibfnamefont {C.-Y.}\ \bibnamefont {Lu}}, \bibinfo
  {author} {\bibfnamefont {H.}~\bibnamefont {Weinfurter}}, \bibinfo {author}
  {\bibfnamefont {A.}~\bibnamefont {Zeilinger}}, \ and\ \bibinfo {author}
  {\bibfnamefont {M.}~\bibnamefont {\.{Z}ukowski}},\ }\href@noop {} {\bibfield
  {journal} {\bibinfo  {journal} {\rmp}\ }\textbf {\bibinfo {volume} {84}},\
  \bibinfo {pages} {777} (\bibinfo {year} {2012})}\BibitemShut {NoStop}%
\bibitem [{\citenamefont {Einstein}\ \emph {et~al.}(1935)\citenamefont
  {Einstein}, \citenamefont {Podolsky},\ and\ \citenamefont {Rosen}}]{EPR35}%
  \BibitemOpen
  \bibfield  {author} {\bibinfo {author} {\bibfnamefont {A.}~\bibnamefont
  {Einstein}}, \bibinfo {author} {\bibfnamefont {B.}~\bibnamefont {Podolsky}},
  \ and\ \bibinfo {author} {\bibfnamefont {N.}~\bibnamefont {Rosen}},\
  }\href@noop {} {\bibfield  {journal} {\bibinfo  {journal} {Phys, Rev.}\
  }\textbf {\bibinfo {volume} {47}},\ \bibinfo {pages} {777} (\bibinfo {year}
  {1935})}\BibitemShut {NoStop}%
\bibitem [{\citenamefont {Mermin}(1990)}]{Mermin90c}%
  \BibitemOpen
  \bibfield  {author} {\bibinfo {author} {\bibfnamefont {N.~D.}\ \bibnamefont
  {Mermin}},\ }\href@noop {} {\bibfield  {journal} {\bibinfo  {journal} {Phys.
  Rev. Lett.}\ }\textbf {\bibinfo {volume} {65}},\ \bibinfo {pages} {1838}
  (\bibinfo {year} {1990})}\BibitemShut {NoStop}%
\bibitem [{\citenamefont {Werner}\ and\ \citenamefont {Wolf}(2001)}]{Werner01}%
  \BibitemOpen
  \bibfield  {author} {\bibinfo {author} {\bibfnamefont {R.~F.}\ \bibnamefont
  {Werner}}\ and\ \bibinfo {author} {\bibfnamefont {M.~M.}\ \bibnamefont
  {Wolf}},\ }\href@noop {} {\bibfield  {journal} {\bibinfo  {journal} {Phys.
  Rev. A}\ }\textbf {\bibinfo {volume} {64}},\ \bibinfo {pages} {032112}
  (\bibinfo {year} {2001})}\BibitemShut {NoStop}%
\bibitem [{\citenamefont {\.{Z}ukowski}\ and\ \citenamefont {\v{C}.
  Brukner}(2002)}]{Zukowski02}%
  \BibitemOpen
  \bibfield  {author} {\bibinfo {author} {\bibfnamefont {M.}~\bibnamefont
  {\.{Z}ukowski}}\ and\ \bibinfo {author} {\bibnamefont {\v{C}. Brukner}},\
  }\href@noop {} {\bibfield  {journal} {\bibinfo  {journal} {Phys. Rev. Lett.}\
  }\textbf {\bibinfo {volume} {88}},\ \bibinfo {pages} {210401} (\bibinfo
  {year} {2002})}\BibitemShut {NoStop}%
\bibitem [{\citenamefont {Collins}\ \emph {et~al.}(2002)\citenamefont
  {Collins}, \citenamefont {Gisin}, \citenamefont {Linden}, \citenamefont
  {Massar},\ and\ \citenamefont {Popescu}}]{Collins02}%
  \BibitemOpen
  \bibfield  {author} {\bibinfo {author} {\bibfnamefont {D.}~\bibnamefont
  {Collins}}, \bibinfo {author} {\bibfnamefont {N.}~\bibnamefont {Gisin}},
  \bibinfo {author} {\bibfnamefont {N.}~\bibnamefont {Linden}}, \bibinfo
  {author} {\bibfnamefont {S.}~\bibnamefont {Massar}}, \ and\ \bibinfo {author}
  {\bibfnamefont {S.}~\bibnamefont {Popescu}},\ }\href@noop {} {\bibfield
  {journal} {\bibinfo  {journal} {Phys. Rev. Lett.}\ }\textbf {\bibinfo
  {volume} {88}},\ \bibinfo {pages} {040404} (\bibinfo {year}
  {2002})}\BibitemShut {NoStop}%
\bibitem [{\citenamefont {Laskowski}\ \emph {et~al.}(2004)\citenamefont
  {Laskowski}, \citenamefont {Paterek}, \citenamefont {\.{Z}ukowski},\ and\
  \citenamefont {\v{C}. Brukner}}]{Laskowski04}%
  \BibitemOpen
  \bibfield  {author} {\bibinfo {author} {\bibfnamefont {W.}~\bibnamefont
  {Laskowski}}, \bibinfo {author} {\bibfnamefont {T.}~\bibnamefont {Paterek}},
  \bibinfo {author} {\bibfnamefont {M.}~\bibnamefont {\.{Z}ukowski}}, \ and\
  \bibinfo {author} {\bibnamefont {\v{C}. Brukner}},\ }\href@noop {} {\bibfield
   {journal} {\bibinfo  {journal} {Phys. Rev. Lett.}\ }\textbf {\bibinfo
  {volume} {93}},\ \bibinfo {pages} {200401} (\bibinfo {year}
  {2004})}\BibitemShut {NoStop}%
\bibitem [{\citenamefont {Son}\ \emph {et~al.}(2006)\citenamefont {Son},
  \citenamefont {Lee},\ and\ \citenamefont {Kim}}]{Son06}%
  \BibitemOpen
  \bibfield  {author} {\bibinfo {author} {\bibfnamefont {W.}~\bibnamefont
  {Son}}, \bibinfo {author} {\bibfnamefont {J.}~\bibnamefont {Lee}}, \ and\
  \bibinfo {author} {\bibfnamefont {M.~S.}\ \bibnamefont {Kim}},\ }\href@noop
  {} {\bibfield  {journal} {\bibinfo  {journal} {Phys. Rev. Lett.}\ }\textbf
  {\bibinfo {volume} {96}},\ \bibinfo {pages} {060406} (\bibinfo {year}
  {2006})}\BibitemShut {NoStop}%
\bibitem [{\citenamefont {Lim}\ \emph {et~al.}(2010)\citenamefont {Lim},
  \citenamefont {Ryu}, \citenamefont {Yoo}, \citenamefont {Lee}, \citenamefont
  {Bang},\ and\ \citenamefont {Lee}}]{James10}%
  \BibitemOpen
  \bibfield  {author} {\bibinfo {author} {\bibfnamefont {J.}~\bibnamefont
  {Lim}}, \bibinfo {author} {\bibfnamefont {J.}~\bibnamefont {Ryu}}, \bibinfo
  {author} {\bibfnamefont {S.}~\bibnamefont {Yoo}}, \bibinfo {author}
  {\bibfnamefont {C.}~\bibnamefont {Lee}}, \bibinfo {author} {\bibfnamefont
  {J.}~\bibnamefont {Bang}}, \ and\ \bibinfo {author} {\bibfnamefont
  {J.}~\bibnamefont {Lee}},\ }\href@noop {} {\bibfield  {journal} {\bibinfo
  {journal} {New J. Phys.}\ }\textbf {\bibinfo {volume} {12}},\ \bibinfo
  {pages} {103012} (\bibinfo {year} {2010})}\BibitemShut {NoStop}%
\bibitem [{\citenamefont {Ekert}(1991)}]{Ekert91}%
  \BibitemOpen
  \bibfield  {author} {\bibinfo {author} {\bibfnamefont {A.~K.}\ \bibnamefont
  {Ekert}},\ }\href@noop {} {\bibfield  {journal} {\bibinfo  {journal} {Phys.
  Rev. Lett.}\ }\textbf {\bibinfo {volume} {67}},\ \bibinfo {pages} {661}
  (\bibinfo {year} {1991})}\BibitemShut {NoStop}%
\bibitem [{\citenamefont {Horodecki}\ \emph {et~al.}(1996)\citenamefont
  {Horodecki}, \citenamefont {Horodecki},\ and\ \citenamefont
  {Horodecki}}]{Horodecki96}%
  \BibitemOpen
  \bibfield  {author} {\bibinfo {author} {\bibfnamefont {R.}~\bibnamefont
  {Horodecki}}, \bibinfo {author} {\bibfnamefont {M.}~\bibnamefont
  {Horodecki}}, \ and\ \bibinfo {author} {\bibfnamefont {P.}~\bibnamefont
  {Horodecki}},\ }\href@noop {} {\bibfield  {journal} {\bibinfo  {journal}
  {Phys. Lett. A}\ }\textbf {\bibinfo {volume} {222}},\ \bibinfo {pages} {21}
  (\bibinfo {year} {1996})}\BibitemShut {NoStop}%
\bibitem [{\citenamefont {Cleve}\ and\ \citenamefont
  {Buhrman}(1997)}]{Cleve97}%
  \BibitemOpen
  \bibfield  {author} {\bibinfo {author} {\bibfnamefont {R.}~\bibnamefont
  {Cleve}}\ and\ \bibinfo {author} {\bibfnamefont {H.}~\bibnamefont
  {Buhrman}},\ }\href@noop {} {\bibfield  {journal} {\bibinfo  {journal} {Phys.
  Rev. A}\ }\textbf {\bibinfo {volume} {56}},\ \bibinfo {pages} {1201}
  (\bibinfo {year} {1997})}\BibitemShut {NoStop}%
\bibitem [{\citenamefont {\v{C}. Brukner}\ \emph {et~al.}(2004)\citenamefont
  {\v{C}. Brukner}, \citenamefont {\.{Z}ukowski}, \citenamefont {Pan},\ and\
  \citenamefont {Zeilinger}}]{Brukner04}%
  \BibitemOpen
  \bibfield  {author} {\bibinfo {author} {\bibnamefont {\v{C}. Brukner}},
  \bibinfo {author} {\bibfnamefont {M.}~\bibnamefont {\.{Z}ukowski}}, \bibinfo
  {author} {\bibfnamefont {J.-W.}\ \bibnamefont {Pan}}, \ and\ \bibinfo
  {author} {\bibfnamefont {A.}~\bibnamefont {Zeilinger}},\ }\href@noop {}
  {\bibfield  {journal} {\bibinfo  {journal} {Phys. Rev. Lett.}\ }\textbf
  {\bibinfo {volume} {92}},\ \bibinfo {pages} {127901} (\bibinfo {year}
  {2004})}\BibitemShut {NoStop}%
\bibitem [{\citenamefont {\.{Z}ukowski}\ \emph {et~al.}(1998)\citenamefont
  {\.{Z}ukowski}, \citenamefont {Zeilinger}, \citenamefont {Horne},\ and\
  \citenamefont {Weinfurter}}]{Zukowski98}%
  \BibitemOpen
  \bibfield  {author} {\bibinfo {author} {\bibfnamefont {M.}~\bibnamefont
  {\.{Z}ukowski}}, \bibinfo {author} {\bibfnamefont {A.}~\bibnamefont
  {Zeilinger}}, \bibinfo {author} {\bibfnamefont {M.~A.}\ \bibnamefont
  {Horne}}, \ and\ \bibinfo {author} {\bibfnamefont {H.}~\bibnamefont
  {Weinfurter}},\ }\href@noop {} {\bibfield  {journal} {\bibinfo  {journal}
  {Acta Phys. Pol. A}\ }\textbf {\bibinfo {volume} {93}},\ \bibinfo {pages}
  {187} (\bibinfo {year} {1998})}\BibitemShut {NoStop}%
\bibitem [{\citenamefont {Hillery}\ \emph {et~al.}(1999)\citenamefont
  {Hillery}, \citenamefont {Bu\v{z}ek},\ and\ \citenamefont
  {Berthiaume}}]{Hillery99}%
  \BibitemOpen
  \bibfield  {author} {\bibinfo {author} {\bibfnamefont {M.}~\bibnamefont
  {Hillery}}, \bibinfo {author} {\bibfnamefont {V.}~\bibnamefont {Bu\v{z}ek}},
  \ and\ \bibinfo {author} {\bibfnamefont {A.}~\bibnamefont {Berthiaume}},\
  }\href@noop {} {\bibfield  {journal} {\bibinfo  {journal} {Phys. Rev. A}\
  }\textbf {\bibinfo {volume} {59}},\ \bibinfo {pages} {1829} (\bibinfo {year}
  {1999})}\BibitemShut {NoStop}%
\bibitem [{\citenamefont {Kempe}(1999)}]{Kempe99}%
  \BibitemOpen
  \bibfield  {author} {\bibinfo {author} {\bibfnamefont {J.}~\bibnamefont
  {Kempe}},\ }\href@noop {} {\bibfield  {journal} {\bibinfo  {journal} {Phys.
  Rev. A}\ }\textbf {\bibinfo {volume} {60}},\ \bibinfo {pages} {910} (\bibinfo
  {year} {1999})}\BibitemShut {NoStop}%
\bibitem [{\citenamefont {Scarani}\ and\ \citenamefont
  {Gisin}(2001)}]{Scarani01}%
  \BibitemOpen
  \bibfield  {author} {\bibinfo {author} {\bibfnamefont {V.}~\bibnamefont
  {Scarani}}\ and\ \bibinfo {author} {\bibfnamefont {N.}~\bibnamefont
  {Gisin}},\ }\href@noop {} {\bibfield  {journal} {\bibinfo  {journal} {Phys.
  Rev. Lett.}\ }\textbf {\bibinfo {volume} {87}},\ \bibinfo {pages} {117901}
  (\bibinfo {year} {2001})}\BibitemShut {NoStop}%
\bibitem [{\citenamefont {Barrett}\ \emph {et~al.}(2005)\citenamefont
  {Barrett}, \citenamefont {Hardy},\ and\ \citenamefont {Kent}}]{Barrett05}%
  \BibitemOpen
  \bibfield  {author} {\bibinfo {author} {\bibfnamefont {J.}~\bibnamefont
  {Barrett}}, \bibinfo {author} {\bibfnamefont {L.}~\bibnamefont {Hardy}}, \
  and\ \bibinfo {author} {\bibfnamefont {A.}~\bibnamefont {Kent}},\ }\href@noop
  {} {\bibfield  {journal} {\bibinfo  {journal} {Phys. Rev. Lett.}\ }\textbf
  {\bibinfo {volume} {95}},\ \bibinfo {pages} {010503} (\bibinfo {year}
  {2005})}\BibitemShut {NoStop}%
\bibitem [{\citenamefont {Ac\'{i}n}\ \emph {et~al.}(2007)\citenamefont
  {Ac\'{i}n}, \citenamefont {Brunner}, \citenamefont {Gisin}, \citenamefont
  {Massar}, \citenamefont {Pironio},\ and\ \citenamefont {Scarani}}]{Acin07}%
  \BibitemOpen
  \bibfield  {author} {\bibinfo {author} {\bibfnamefont {A.}~\bibnamefont
  {Ac\'{i}n}}, \bibinfo {author} {\bibfnamefont {N.}~\bibnamefont {Brunner}},
  \bibinfo {author} {\bibfnamefont {N.}~\bibnamefont {Gisin}}, \bibinfo
  {author} {\bibfnamefont {S.}~\bibnamefont {Massar}}, \bibinfo {author}
  {\bibfnamefont {S.}~\bibnamefont {Pironio}}, \ and\ \bibinfo {author}
  {\bibfnamefont {V.}~\bibnamefont {Scarani}},\ }\href@noop {} {\bibfield
  {journal} {\bibinfo  {journal} {Phys. Rev. Lett.}\ }\textbf {\bibinfo
  {volume} {98}},\ \bibinfo {pages} {230501} (\bibinfo {year}
  {2007})}\BibitemShut {NoStop}%
\bibitem [{\citenamefont {Pironio}\ \emph {et~al.}(2010)\citenamefont
  {Pironio}, \citenamefont {Ac\'{i}n}, \citenamefont {Massar}, \citenamefont
  {de~la Giroday}, \citenamefont {Matsukevich}, \citenamefont {Maunz},
  \citenamefont {Olmschenk}, \citenamefont {Hayes}, \citenamefont {Luo},
  \citenamefont {Manning},\ and\ \citenamefont {Monroe}}]{Pironio10}%
  \BibitemOpen
  \bibfield  {author} {\bibinfo {author} {\bibfnamefont {S.}~\bibnamefont
  {Pironio}}, \bibinfo {author} {\bibfnamefont {A.}~\bibnamefont {Ac\'{i}n}},
  \bibinfo {author} {\bibfnamefont {S.}~\bibnamefont {Massar}}, \bibinfo
  {author} {\bibfnamefont {A.~B.}\ \bibnamefont {de~la Giroday}}, \bibinfo
  {author} {\bibfnamefont {D.~N.}\ \bibnamefont {Matsukevich}}, \bibinfo
  {author} {\bibfnamefont {P.}~\bibnamefont {Maunz}}, \bibinfo {author}
  {\bibfnamefont {S.}~\bibnamefont {Olmschenk}}, \bibinfo {author}
  {\bibfnamefont {D.}~\bibnamefont {Hayes}}, \bibinfo {author} {\bibfnamefont
  {L.}~\bibnamefont {Luo}}, \bibinfo {author} {\bibfnamefont {T.~A.}\
  \bibnamefont {Manning}}, \ and\ \bibinfo {author} {\bibfnamefont
  {C.}~\bibnamefont {Monroe}},\ }\href@noop {} {\bibfield  {journal} {\bibinfo
  {journal} {Nature}\ }\textbf {\bibinfo {volume} {464}},\ \bibinfo {pages}
  {1021} (\bibinfo {year} {2010})}\BibitemShut {NoStop}%
\bibitem [{\citenamefont {\.{Z}ukowski}\ and\ \citenamefont
  {Kaszlikowski}(1999)}]{Zukowski99}%
  \BibitemOpen
  \bibfield  {author} {\bibinfo {author} {\bibfnamefont {M.}~\bibnamefont
  {\.{Z}ukowski}}\ and\ \bibinfo {author} {\bibfnamefont {D.}~\bibnamefont
  {Kaszlikowski}},\ }\href@noop {} {\bibfield  {journal} {\bibinfo  {journal}
  {Phys. Rev. A}\ }\textbf {\bibinfo {volume} {59}},\ \bibinfo {pages} {3200}
  (\bibinfo {year} {1999})}\BibitemShut {NoStop}%
\bibitem [{\citenamefont {Kaszlikowski}\ and\ \citenamefont
  {\.{Z}ukowski}(2002)}]{Kaszlikowski02}%
  \BibitemOpen
  \bibfield  {author} {\bibinfo {author} {\bibfnamefont {D.}~\bibnamefont
  {Kaszlikowski}}\ and\ \bibinfo {author} {\bibfnamefont {M.}~\bibnamefont
  {\.{Z}ukowski}},\ }\href@noop {} {\bibfield  {journal} {\bibinfo  {journal}
  {Phys. Rev. A}\ }\textbf {\bibinfo {volume} {66}},\ \bibinfo {pages} {042107}
  (\bibinfo {year} {2002})}\BibitemShut {NoStop}%
\bibitem [{\citenamefont {Cerf}\ \emph {et~al.}(2002)\citenamefont {Cerf},
  \citenamefont {Massar},\ and\ \citenamefont {Pironio}}]{Cerf02a}%
  \BibitemOpen
  \bibfield  {author} {\bibinfo {author} {\bibfnamefont {N.~J.}\ \bibnamefont
  {Cerf}}, \bibinfo {author} {\bibfnamefont {S.}~\bibnamefont {Massar}}, \ and\
  \bibinfo {author} {\bibfnamefont {S.}~\bibnamefont {Pironio}},\ }\href@noop
  {} {\bibfield  {journal} {\bibinfo  {journal} {Phys. Rev. Lett.}\ }\textbf
  {\bibinfo {volume} {89}},\ \bibinfo {pages} {080402} (\bibinfo {year}
  {2002})}\BibitemShut {NoStop}%
\bibitem [{\citenamefont {Lee}\ \emph {et~al.}(2006)\citenamefont {Lee},
  \citenamefont {Lee},\ and\ \citenamefont {Kim}}]{Lee06}%
  \BibitemOpen
  \bibfield  {author} {\bibinfo {author} {\bibfnamefont {J.}~\bibnamefont
  {Lee}}, \bibinfo {author} {\bibfnamefont {S.-W.}\ \bibnamefont {Lee}}, \ and\
  \bibinfo {author} {\bibfnamefont {M.~S.}\ \bibnamefont {Kim}},\ }\href@noop
  {} {\bibfield  {journal} {\bibinfo  {journal} {Phys. Rev. A}\ }\textbf
  {\bibinfo {volume} {73}},\ \bibinfo {pages} {032316} (\bibinfo {year}
  {2006})}\BibitemShut {NoStop}%
\bibitem [{\citenamefont {Tang}\ \emph {et~al.}(2013)\citenamefont {Tang},
  \citenamefont {Yu},\ and\ \citenamefont {Oh}}]{Tang13}%
  \BibitemOpen
  \bibfield  {author} {\bibinfo {author} {\bibfnamefont {W.}~\bibnamefont
  {Tang}}, \bibinfo {author} {\bibfnamefont {S.}~\bibnamefont {Yu}}, \ and\
  \bibinfo {author} {\bibfnamefont {C.~H.}\ \bibnamefont {Oh}},\ }\href@noop {}
  {\bibfield  {journal} {\bibinfo  {journal} {Phys. Rev. Lett.}\ }\textbf
  {\bibinfo {volume} {110}},\ \bibinfo {pages} {100403} (\bibinfo {year}
  {2013})}\BibitemShut {NoStop}%
\bibitem [{Note1()}]{Note1}%
  \BibitemOpen
  \bibinfo {note} {Note that compatible observables are clearly
  concurrent.}\BibitemShut {Stop}%
\bibitem [{Note2()}]{Note2}%
  \BibitemOpen
  \bibinfo {note} {Here, we shall use local unitary observables $\protect
  \mathaccentV {hat}05E{V}=\DOTSB \sum@ \slimits@ _{n=0}^{D-1}\omega ^n \left
  |n\right >_{\protect \mathrm {v}}\left <n\right |$, where $\omega =\protect
  \qopname \relax o{exp}(2 \pi i/D)$, in $D$-dimensional Hilbert space. The
  observable $\protect \mathaccentV {hat}05E{V}$ is unitary, however, one can
  uniquely relate it with a Hermitian observable $\protect \mathaccentV
  {hat}05E{H}$ by requiring $\protect \mathaccentV {hat}05E{V}=\protect
  \qopname \relax o{exp}(i \protect \mathaccentV {hat}05E{H})$. Therefore the
  complex eigenvalues of $\protect \mathaccentV {hat}05E{V}$ can be associated
  with the measurement results, denoted by real eigenvalues of $\protect
  \mathaccentV {hat}05E{H}$. Such a unitary representation leads to a
  simplification of mathematics without changing any physical results~\cite
  {Cerf02a, Lee06}.}\BibitemShut {Stop}%
\end{thebibliography}%

\end{document}